\documentclass[aps,twocolumn,showpacs]{revtex4}
\usepackage{epsfig}
\usepackage{graphicx}
\usepackage{amsfonts}
\usepackage[figuresright]{rotating} 
\usepackage{amssymb}
\usepackage{amsmath}
\usepackage{psfrag}

\def\avg#1{\langle#1\rangle}
\def\Re{\rm{Re}}

\def\be{\begin{equation}} \def\ee{\end{equation}}
\def\bea{\begin{eqnarray}} \def\eea{\end{eqnarray}}

\def\nn{\nonumber}

\def\Re{\mbox{Re}}

\begin{document}

\title{Mixed triplet and singlet pairing in ultracold multicomponent fermion 
systems with dipolar interactions}
\author{Congjun Wu}
\affiliation{Department of Physics, University of California, San Diego,
CA 92093 }

\author{J. E. Hirsch}
\affiliation{Department of Physics, University of California, San Diego,
CA 92093 }

\begin{abstract}
The symmetry properties of the Cooper pairing problem  for 
multi-component ultra-cold dipolar molecular systems are investigated.
The dipolar anisotropy provides a natural and robust mechanism for both 
triplet and singlet Cooper pairing to first order in the interaction 
strength.  
With a purely dipolar interaction, the triplet $p_z$-like polar  pairing is 
the most dominant.
A short-range attractive interaction can enhance the singlet
pairing to be nearly degenerate with the triplet pairing.
We point out that these two pairing channels can mix by developing 
a relative phase of 
$\pm\frac{\pi}{2}$, thus spontaneously breaking time-reversal symmetry. 
We also suggest the possibility of such mixing of triplet and singlet 
pairing in other systems. 
\end{abstract}
 

\maketitle

The study of ultracold dipolar molecules has recently become 
a research focus of cold atom physics
\cite{ospelkaus2008,ni2008,griesmaier2005,mcclelland2006}.
The prominent feature of the dipolar interaction is its
$d_{r^2-3z^2}$-type anisotropy when the dipolar 
moments are aligned by an external electric field.
Considerable progress has been made in studying anisotropic 
condensation of  dipolar bosons 
\cite{koch2008,lahaye2009,lahaye2009a,menotti2007a}.
Furthermore, dipolar fermionic systems provide an exciting opportunity 
to study exotic anisotropic many-body physics of fermions.
Experimentally, a near quantum-degenerate gas of the dipolar fermion
$^{40}$K-$^{87}$Rb has been achieved \cite{ospelkaus2008}. 
A number of theoretical works have been done for the anisotropic Fermi
liquid properties of dipolar Fermi gases \cite{sogo2008,miyakawa2008,ronen2009,
chan2009,fregoso2009}, including both singlet particle
and collective excitations.

The dipolar interaction also has important effects in the Cooper
pairing symmetry as studied in Refs \cite{baranov2002,baranov2004,
baranov2008a,you1999,bruun2008}.
In the single component case, the only possible pairing channels
are of odd parity.
Assuming dipole moments along the $z$-axis, 
the pairing symmetry is mainly of $p_z$
with slight hybridization with other odd partial wave components.
Dipolar molecules can have an internal degree of freedom arising from 
the hyperfine configurations of the constituent atoms.
The electric dipolar interaction is independent of these internal
components which will be denoted as spin below.
The inter-component interaction opens up the possibility of both spin 
singlet and triplet pairings for the simplest two-component case.
It would be interesting to study even richer 
Cooper pairing patterns and the competition among them.
 
In this article, we show that the dipolar interaction  favors
Cooper pairing  in the triplet channel  over  the singlet channel.
This is an effect directly arising from the anisotropy of
the dipolar interaction, and it occurs to  first order in  the
interaction strength.
In contrast, it does not appear in the usual condensed 
matter triplet pairing systems such as  superfluid 
$^3$He \cite{leggett1975,VOLLHARDT1990,bw}:
the spin fluctuation mechanism based on the strong ferromagnetic
tendency in $^3$He arises from the repulsive part of the $^3$He-$^3$He 
interaction at second order.
For a two-component dipolar fermion system, we find 
the dominate pairing in the spin triplet $p_z$-like channel
with the purely dipolar interaction.
It can mix with the singlet $s+d_{r^2-3z^2}$-pairing whose
pairing strength is tunable through the 
short range non-dipolar $s$-wave scattering.
The mixing occurs with a relative phase of $\pm\frac{\pi}{2}$ 
which breaks time-reversal (TR) symmetry  spontaneously.
Pairing in dipolar Fermi gases with more than two components
is also discussed. 

Samokhin {\it et al.} studied non-uniform mixed parity 
superfluid states in the presence of dipolar interactions \cite{samokhin2006}.
They considered coupling 
between singlet and triplet  channels  with zero relative phase only. 
Kabanov \cite{kabanov} also considered recently mixture of 
singlet and triplet pairing with zero relative phase.
While this work was being completed, a study of the competition 
between triplet and singlet pairing in dipolar fermionic systems 
appeared that analyzed some of the cases considered here \cite{shi}. 

We begin with the two-component dipolar fermionic system  with
the electric dipole moments aligned along the $z$-axis.
The dipolar interaction reads
$V_{3D}(\vec r_1-\vec r_2)= -{2 d^2 \over {|\vec r_1-\vec r_2|^3}} 
P_2 (\cos \theta _{\vec r_1-\vec r_2})$,
where $\theta_{\vec r_1-\vec r_2}$ is the angle between ($\vec r_1 -\vec r_2$)
and the electric field $\vec E$; $d$ is the electric dipole moment. 
The anisotropy
is manifested  in the angular dependence of the $V_{3D}$ with the form
of the second Legendre polynomial.
The Fourier transform of the 3D dipolar interaction, $V(\vec k)=
\frac{8\pi d^2}{3} P_2(\cos\theta_k)$, only depends on the polar angle 
of $\vec k$.
The Hamiltonian is written as
\bea
H&=& \sum_{k,\alpha} \big[\epsilon(\vec k)-\mu \big ]
c^\dagger_{\alpha}(\vec k) c_{\alpha} (\vec k)  +
\frac{1}{2V} \sum_{k,k^\prime,q} V(\vec k-\vec k^\prime)\nn \\
&\times& P^\dagger_{\beta\alpha} (\vec k;\vec q)  P_{\alpha\beta} (\vec k^\prime
;\vec q)
\label{eq:ham}
\eea
where $\epsilon(\vec k)=\hbar k^2/(2m)$; $\mu$ is the chemical
potential; $P_{\alpha\beta} (\vec k;\vec q)
=c_{\alpha}(-\vec k+\vec q) c_{\beta}(\vec k+\vec q)$ is
the pairing operator; $\alpha,\beta$ refer to $\uparrow$ and 
$\downarrow$.
Please note that $V(\vec k-\vec k^\prime)$ depends on the polar angle of the 
vector $\vec k-\vec k^\prime$, not the relative angle between 
$\vec k$ and $\vec k^\prime$.
We define a dimensionless parameter   to describe the interaction strength
as the ratio between the characteristic interaction energy and the
Fermi energy:
$\lambda \equiv E_{int}/E_F = \frac{2}{3} \frac{d^2 m k_f }{\pi^2\hbar^2}$.

We only consider uniform pairing states at the mean-field
level, thus set $\vec q=0$ 
in the pairing interaction in Eq. \ref{eq:ham}.
We
define $P_{\alpha\beta}(\vec k)=P_{\alpha\beta}(\vec k; \vec q=0)$
which satisfies $P_{\alpha\beta}(\vec k)=-P_{\beta\alpha}(-\vec k)$.
The pairing operators can be decomposed into the spin singlet $P_{si}$
and triplet channels $P_{tri}^{x,y,z}$:  
$P_{si}(\vec k)= \frac{1}{\sqrt 2} 
[P_{\uparrow\downarrow}(\vec k)-P_{\downarrow\uparrow}(\vec k)]$,
$P_{tri}^z(\vec k)= \frac{1}{\sqrt 2} 
[P_{\uparrow\downarrow}(\vec k)+P_{\downarrow\uparrow}(\vec k)]$,
$P_{tri}^x(\vec k)= -\frac{1}{\sqrt 2} 
[P_{\uparrow\uparrow}(\vec k)-P_{\downarrow\downarrow}(\vec k)]$,
and $P_{tri}^y(\vec k)= -\frac{i}{\sqrt 2} 
[P_{\uparrow\uparrow}(\vec k)+P_{\downarrow\downarrow}(\vec k)]$.
$P_{si}(\vec k)$ and $P_{tri}^\mu (\vec k)$ are even and odd functions of 
$\vec k$  respectively;
$P_{tri}^\mu$ describes the triplet pairing operators whose total spin
is the eigenstate of $\hat e_\mu \cdot \vec S_{pair}$ with zero eigenvalue.
Using these operators, the pairing interaction of 
Eq. \ref{eq:ham} with $\vec q=0$ can be rewritten as
\bea
H_{pair} &=& \frac{1}{2V}\sum_{k,k^\prime,} \big\{ ~V_{tri}(\vec k;\vec k^\prime)
~[~\sum_{\mu=x,y,z} P^{\dagger,\mu}_{tri}(\vec k) P_{tri}^\mu(\vec k^\prime)~]
\nn \\
&+& V_{si}(\vec k;\vec k^\prime) P^\dagger_{si}(\vec k) P_{si}(\vec k^\prime)~\big\},
\label{Eq:polarpair}
\eea
where $V_{tri,si}(\vec k;\vec k^\prime)=\frac{1}{2} \big\{ V(\vec k-\vec k^\prime)
\mp V(\vec k +\vec k^\prime)\big\}$.
$V_{si}(\vec k;\vec k^\prime)$ is an even function of both arguments
 $\vec k$ and $\vec k^\prime$,
while $V_{tri}(\vec k;\vec k^\prime)$ is an odd function of both.

The decoupled mean-field Hamiltonian reads
\bea
H_{mf}&=&\sum^\prime_{\vec k} \Psi^\dagger (\vec k) 
\left(\begin{array}{cc}
\xi(\vec k) I & \Delta_{\alpha\beta} (\vec k) \\
\Delta^*_{\beta\alpha}(\vec k) &-\xi(\vec k) I
\end{array}
\right) \Psi(\vec k),
\eea
where  $\sum_k^\prime$ means summation over half of momentum space;
$\xi (\vec k)=\epsilon (\vec k) -\mu$;
$\Psi(\vec k)=(c_\uparrow (\vec k), c_\downarrow(\vec k),
c_\uparrow^\dagger(-\vec k), c_\downarrow^\dagger(-\vec k) )^T$.
The mean-field gap function is defined as
$ \Delta_{\alpha\beta}(\vec k)=\frac{1}{V} \sum_{\vec k^\prime}
V(\vec k-\vec k^\prime) \avg{|P_{\alpha\beta}(\vec k^\prime)|}$.
$\Delta_{\alpha\beta}$ can be decomposed into   singlet and triplet channels as 
$\Delta_{\alpha\beta}(\vec k)=\Delta_{si} (\vec k) i\sigma^y_{\alpha\beta}
+\Delta_{tri,\mu} (\vec k) (i\sigma^\mu \sigma^y)_{\alpha\beta} $.
The Bogoliubov quasiparticle spectra become
$E_{1,2}(\vec k)=\sqrt{\xi_k^2+\lambda^2_{1,2}(\vec k) }$ where
$\lambda^2_{1,2}(\vec k)$ are the eigenvalues of the positive-definite
Hermitian matrix   $\Delta^\dagger(\vec k) \Delta(\vec k)$.
Its trace satisfies
$\lambda^2_1(\vec k)+\lambda^2_2(\vec k)=
|\Delta_{si}(\vec k)|^2+\sum_\mu |\Delta_{tri,\mu} (\vec k)|^2$.
The free energy becomes
\bea
F&=&-\frac{2}{\beta}\sum_{k,i=1,2} \ln \big[2\cosh 
\frac{\beta E_{\vec k,i} }{2}\big]
-\frac{1}{2V} \sum_{\vec k,\vec k^\prime,a=si, (tri,\mu)} \nn \\
&\times& \big \{ \Delta^*_{a}(\vec k) V^{-1}_{a}(\vec k;\vec k^\prime) 
\Delta_a(\vec k)
\big\},
\label{eq:free}
\eea
where $V^{-1}_{si,tri}(\vec k;\vec k^\prime)$ is the inverse of the interaction
matrix defined as $\frac{1}{V}\sum_{k^\prime} V_{si,tri}(\vec k, \vec k^\prime)
V^{-1}_{si,tri}(\vec k^\prime, \vec k^{\prime\prime})
=\delta_{\vec k, \vec k^{\prime\prime}}$.

\begin{figure}
\centering\epsfig{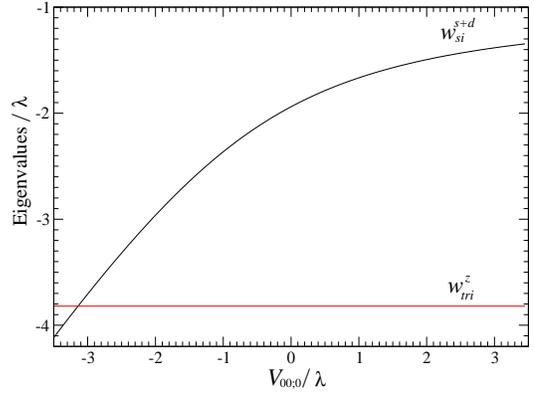}
\caption{(Color online) The eigenvalues $w_{tri}^{z}$ for the spin triplet
$\phi^{z}$-channel and $w_{si}^{s+d}$ for the singlet $\phi^{s+d}$-channel.
The latter depends on the short-range $s$-wave interaction
$V_{00;0}$.
}
\label{fig:dipolar}
\end{figure}

The gap equations are expressed as
\bea
\Delta_{tri,\mu}(\vec k)&=&-\int \frac{d^3 k^\prime}{(2\pi)^3}
V_{tri}(\vec k;\vec k^\prime) K(\vec k^\prime)
\Delta_{tri,\mu} (\vec k^\prime), \nn \\
\Delta_{si}(\vec k)&=&-\int \frac{d^3 k^\prime}{(2\pi)^3}
V_{si}(\vec k;\vec k^\prime) K(\vec k^\prime)
\Delta_{si} (\vec k^\prime),
\label{eq:gap}
\eea
where $K(\vec k^\prime)=\tanh[\frac{\beta}{2} 
E_{i}(\vec k^\prime)]/[2E_i (\vec k^\prime)]$.
Eq. \ref{eq:gap} formally diverges.
It can be regularized following the standard procedure explained
in Ref. \cite{baranov2002,gurarie2007} by replacing the bare interaction
$V_{tri,si}$ with the renormalized zero energy vertex functions $\Gamma_{tri,si}$.
At the level of the Born approximation, this regularization is 
equivalent to just introducing an energy cutoff of $\pm\bar \omega$ 
for $\xi_{\vec k}^\prime$ in Eq. \ref{eq:gap}, where $\bar\omega$
is at the scale of the Fermi energy.

To analyze the dominant pairing instability around $T_c$, 
Eq. \ref{eq:gap} is linearized.
Considering that the strongest pairing occurs at the  Fermi surface
and following the standard procedure in Ref. \cite{baranov2002},
we define
the eigen-gap functions  $\phi_{tri,si}^j (\vec k)$ 
 satisfying
\bea
N_0 \int \frac{d\Omega_{k^\prime}}{4\pi} V_{tri,si} (\vec k; \vec k^\prime) 
\phi_{tri,si}^j (\vec k^\prime)
=w_{tri,si}^j \phi^j_{tri,si}(\vec k)
\eea
where $N_0=\frac{mk_f}{\pi^2\hbar^2}$ is the density of state at 
the  Fermi surface; $w_{tri,si}^j$ are dimensionless eigenvalues; 
$\vec k,\vec k^\prime$ are at the Fermi surface.
We neglect the effect of Fermi surface distortion on pairing
which is a higher order effect  in  the interaction strength $\lambda$.
Then Eq. \ref{eq:gap} is linearized into a set of decoupled equations
\bea
\phi^j_{tri,si} \{ 1 +w_{tri,si} \ln [(2 e^{\gamma} \bar \omega)/(\pi k_B T)] \}=0.
\label{eq:Tc}
\eea

The spherical harmonics decomposition of  $V_{tri,si} (\vec k;\vec k^\prime)$ 
reads
\bea
\frac{N_0}{4\pi} V_{tri,si}(\vec k;\vec k^\prime)=\sum_{l,l^\prime;m}
V_{ll^\prime;m}
Y_{lm}^*(\Omega_k) Y_{l^\prime m} (\Omega_{\vec k^\prime}),
\label{eq:partialwave}
\eea
where $V_{ll^\prime;m}$ remains diagonal for $m$ but couples partial wave
channels with $l^\prime=l, l\pm 2$.
$V_{tri,si}$ only have matrix elements in odd and even partial wave
channels, respectively.
$V_{ll;m}$ has the same expressions of the Landau parameters of the 
dipolar Fermi gases given by Fregoso {\it et al.} \cite{fregoso2009} 
except for  an overall minus sign due to pairing and a trivial overall numerical
factor difference.
For $l=l^\prime=m=0$, $V_{00;0}=0$ because the average value of the dipolar
interaction is zero.

We diagonalize the matrix $V_{ll^\prime;m}$ to find the dominant 
negative eigenvalues which determine the dominant pairing channels.
Two eigenvectors are found with 
eigenvalues much more negative than other channels.
One lies in the triplet odd parity sector with dominant 
$p_z$-wave character with slight hybridization with other odd
parity channels as same as in the single component case
\cite{baranov2002}: $\phi^z(\Omega_k)$ with the
most negative eigenvalue $w_{tri}^z=-3.820\lambda$,
 whose eigenvector is $\phi^z(\Omega_k)\approx 0.993 Y_{10}-0.120
Y_{30}$.
The other one lies in the even parity spin singlet channel.
For the purely dipolar interaction, its eigenvalue is
$w^{s+d}_{si}=-1.935\lambda$ and the eigenvector
lies in the mixed channel of $s+d_{k^2-3k_z^2}$ 
as $\phi^{s+d}(\Omega_k)\approx 0.6 Y_{00}-0.8 Y_{20}$
with nodes.
However, this channel is sensitive to the strength of the short range $s$-wave 
scattering, which contributes only to the matrix element of $V_{00;0}$
as depicted in Fig. \ref{fig:dipolar}.
Experimentally, this scattering can be tuned to the scale of 
the Fermi energy through Feshbach resonance, {\it i.e.}, 
$V_{00;0}$ can become of order  1. 
As for $\lambda$, as estimated in Ref. \cite{fregoso2009}, it 
could reach $0.1 \sim 0.2$.
Thus the competition between $\phi^{s+d}$ and $\phi^s$ can be studied
experimentally in the future. 
When they become degenerate at $V_{00;0}/\lambda\approx -3.15$,
$\phi^{s+d}$ becomes mostly of $s$-wave character
as $\phi^{s+d}(\Omega_k)\approx 0.901 Y_{00}-0.434 Y_{20}$.

We first consider the case of the dominant triplet pairing, 
whose critical temperature is determined from Eq. \ref{eq:Tc}
as $T_{p_z}\approx (2e^\gamma \bar \omega/\pi) \exp (-1/|w_{tri}^z|)$. 
Its order parameter configuration is
$\Delta_{tri,\mu}(\vec k)=\Delta_{tri,\phi^z}(\Omega_k) \hat d_\mu$.
$\hat d$ is a spin space unit vector.
Without losing generality, it is taken along
the $z$-axis as $\hat d=\hat e_z$.
$ \Delta_{tri,\phi^z}(\Omega_k)= \Delta_{tri} e^{i\gamma} \phi^z(\Omega_k)$
where $\gamma$ is the pairing phase.
This phase has  line-nodes on the equator.
It breaks the $U_c(1)$ gauge, and the spin $SU_s(2)$ symmetries, 
but maintains TR, parity, and a $Z_2$-symmetry of the combined operation
of $\hat d_\mu\rightarrow -\hat d_\mu$ and $\gamma\rightarrow \gamma+\pi$
\cite{zhou2003,salomaa1987}.
Its Goldstone manifold is $G=[U_c(1) \otimes SU_s(2)]/[U_s(1)\otimes Z_2]
=[U_c(1)\otimes S_s^2]/Z_2$.
The corresponding low energy excitations include the phonon and the spin-wave
modes.
It supports two different classes of vortices: the usual integer
vortex of superfluidity, and the half-integer quantum vortex of 
superfluidity combined with a $\pi$-disclination of the $d$-vector.

Next we consider the coexistence of the singlet $\phi^{s+d}$ pairing
and the triplet $\phi^z$ pairing, when they become nearly degenerate.
The order parameter can be chosen as 
$\Delta_{si}(\Omega_k)=\Delta_{0} \phi^{s+d}(\Omega_k)$
and $\Delta_{tri,\mu}(\Omega_k)=\Delta_{z} \phi^{z}(\Omega_k) \delta_{\mu,z}$.
An important observation  is that a relative $\pm\frac{\pi}{2}$-phase 
difference between $\Delta_{si}$ and $\Delta_{tri}$ is favored, 
thus spontaneously breaking TR symmetry. This can be proved as follows:
the quasi-particle spectra reads $E_i=\sqrt{\xi^2+\lambda_i^2}$,
with
$\lambda_{1,2}^2=|\Delta_0 \phi^{s+d}(\Omega_k)|^2
+|\Delta_z \phi^z(\Omega_k)|^2 
\pm 2\Re(\Delta^{*}_z \Delta_0)
\phi^{s+d}(\Omega_k) \phi^z(\Omega_k)$.
The last term vanishes for relative phase $\pm\frac{\pi}{2}$ 
between $\Delta_0$ and $\Delta_z$.
In other words, $\phi^z+i\phi^{s+d}$ is unitary pairing, i.e.,
$\Delta^\dagger(\vec k)\Delta (\vec k)$ is an identity matrix up to a factor, and
$\lambda_1^2=\lambda_2^2=|\Delta_0 \phi^{s+d}(\Omega_k)|^2
+|\Delta_z \phi^z(\Omega_k)|^2|$.
All other phase differences give $\lambda_1\neq \lambda_2$,
thus are non-unitary pairing.
To show that unitary pairing is optimal, we follow the method presented in 
Ref. \cite{cheng2009} to
define the  function $f(x)=-\frac{2}{\beta}\ln [2 \cosh \frac{\beta}{2}
\sqrt{\xi^2_k +x }]$, which satisfies $\frac{d^2}{dx^2} f(x) >0$,
thus $f(\lambda_1^2)+f(\lambda_2^2) \ge 2 f(\frac{\lambda_1^2+\lambda_2^2}{2})$.
Then the first term in Eq. \ref{eq:free} is minimized by the unitary pairing, 
and the second term is degenerate for unitary and non-unitary pairings.
Therefore, $\phi^z+i\phi^{s+d}$ is favored.
This is an exotic fully gapped TR breaking pairing state 
because the nodes of $\phi^{s+d}$ and $\phi^s$ do not coincide.
It also breaks parity but is invariant under the combined 
parity and TR operation.
Its Goldstone manifold for the continuous symmetry breaking
is the same as in the purely triplet-$\phi^z$ pairing phase.

The above analysis can  be recaptured in the Ginzburg-Landau (GL) 
framework.
The bulk pairing order parameters are defined as
$\Delta^{z}_\mu=\sum_k \phi^{z}(\vec k) \Delta_{tri,\mu}(\vec k)$
and $\Delta^{s+d}_\mu=\sum_k \phi^{s+d}(\vec k)\Delta_{si}(\vec k)$.
The GL free energy is constructed as 
\bea
F&=& \alpha_{z}(T) \sum_\mu |\Delta_\mu^z|^2
+\alpha_{s+d}(T) |\Delta^{s+d}|^2 +\beta_z |\Delta_\mu^z|^4
\nn \\
&+& \beta_{s+d} |\Delta^{s+d}|^4
+\gamma_1 \sum_{\mu} |\Delta^{z}_\mu|^2 |\Delta^{s+d}|^2\nn \\
&+&
\gamma_2 \sum_{\mu} \{\Delta^{z*}_\mu \Delta^{z*}_\mu
\Delta^{s+d} \Delta^{s+d}+cc\}, 
\eea
where $\alpha_z=N_0 \ln (\frac{T}{T_{p_z}})$, 
$\alpha_{s+d}=N_0 \ln (\frac{T}{T_{s+d}})$,
and $T_{s+d}$ is defined as 
$T_{s+d}\approx (2e^\gamma \bar \omega/\pi) \exp(-1/|w^{s+d}_{si}|)$;
$\beta_z$ and $\beta_{s+d}$ terms are the standard
quartic terms for the triplet $\phi^{z}$ and singlet 
$\phi^{s+d}$ channels, respectively;
$\gamma_{1,2}$ describe the coupling between the 
$\phi^{z}$ and $\phi^{s+d}$-channels.
Following the analysis on a similar problem in Ref. \cite{lee2009},
we consider the situation where the two channels are nearly degenerate
and $T_{p_z}$ is slightly larger than $T_{s+d}$,
then the triplet pairing develops first at $T_c=T_{p_z}$.
Defining $r=(\gamma_1-2|\gamma_2|)/(2\beta_{z})$,
the condition for the second instability to occur is that 
there exists a lower temperature $T^\prime$ below which
$|\alpha_{s+d}(T^\prime)|>r |\alpha_{p_z}(T^\prime)|$.
This can be satisfied for $r<1$, which results in 
$\frac{T^\prime}{T_{c}}= (\frac{T_{s+d}}{T_{c}})^{\frac{1}{1-r}}$.
The $\pm\frac{\pi}{2}$-phase difference between the triplet
and singlet channel pairing requires that $\gamma_2>0$.
In Refs. \cite{samokhin2006,kabanov},  coupling 
between the singlet and triplet pairings through a linear spatial 
derivative is considered, which leads to spatially non-uniform states.
Due to spin conservation, such a term is not allowed here.

It is natural to further consider competing pairings for even larger 
number of components represented by the internal hyperfine spin 
degrees of freedom, which is an even number. 
The dipolar interaction is independent on them, thus the system
has an $SU(2N)$ symmetry.
The $2N\times 2N$ pairing matrix $\Delta_{\alpha\beta}(\vec k)$ can
be classified as $N(2N+1)$-component symmetric (odd parity)
pairing and $N(2N-1)$-component antisymmetric (even parity) 
pairing, which are generalizations of the triplet and singlet 
channel pairings, respectively.
They can be explicitly constructed as follows.
We define the charge conjugation matrix $R$ as $R_{ij}=(-)^i\delta_{i,2N-i+1}$
and the time-reversal operators $T=RC$ where $C$ is complex conjugation.
For $2N=2$, $R$ reduces to the familiar $-i\sigma_y$.
$R$ satisfies $R^T=-R$ and $R^2=-1$.
On the other hand, any $2N\times 2N$ Hermitian matrix can be expanded
in the basis of the identity matrix and  $4N^2-1$ generators of the 
$SU(2N)$ group.
They can be classified as even and odd under TR transformation.
$N(2N+1)$ of them are TR odd which can 
be constructed as spin-tensor matrices $P_\mu$ with odd rank numbers
(e.g. spin, spin-octupoles, etc.), and $N(2N-1)$ of them are
TR even which can be constructed as spin-tensor matrices $Q_\nu$
with even rank numbers (e.g. the identity matrix, spin quadrupole, etc.).
Using $T P_\mu T^{-1}=-P_\mu$, $T Q_\nu T{^-1}=Q_\nu$, and $R^T=-R$,
it can be shown that the matrices of $P_\mu R$ are symmetric 
and $Q_\nu R $ are antisymmetric, respectively.
Thus we can decompose $\Delta_{\alpha\beta}(\Omega_k)= 
\Delta_{asy,\mu}(\Omega_k) (Q_\mu R)_{\alpha\beta}
+\Delta_{sym,\nu} (\Omega_k) (P_\nu R)_{\alpha\beta}$,
where $\Delta_{asy,\mu} (\Omega_k)~ (\mu=1\sim N(2N-1))$ 
and $\Delta_{sym,\nu}(\Omega_k) ~(\nu=1\sim N (2N+1))$
are even and odd functions of $\Omega_k$, respectively.

The eigenvalue analysis for competing pairing channels is the same
as in Eq. \ref{eq:partialwave} by replacing the triplet (singlet) pairing
with the spin-symmetric (antisymmetric) pairing, respectively.
We next consider the unitary pairing in both spin-symmetric and spin
asymmetric channels, respectively.
A convenient choice for the matrix kernels is that $P_z=\sigma_{z}
\otimes I_N$
for the spin symmetric channel, and $Q=I_{2N}$ where $I_N$ and $I_{2N}$ 
are the identity matrices with $N$ and $2N$-dimensions.
The first one corresponds to the pairing of 
$\sum_{i=1\sim 2N} c^\dagger_i(\vec k) c^\dagger_{2N-i} (-\vec k)$,
while the second corresponds to 
$\sum_{i=1\sim 2N} (-)^{i-1} c^\dagger_i(\vec k) c^\dagger_{2N-i} (-\vec k)$.
In the $\phi^z$-channel pairing states, the spin-$SU(2N)$ symmetry
is broken down into $SU(N)\times SU(N)\times U(1)$, thus it has
$2N^2$ branches of spin-wave Goldstone modes.
The vortex configuration is similar to the $N=1$ case including
the usual integer vortex and the half-quantum vortex combined
with a $\pi$-disclination of spin-texture.
Again for the mixing between pairing in the $\phi^{s+d}$ (spin antisymmetric) 
and $\phi^{z}$ (spin symmetric) channels, a relative phase 
$\pm \frac{\pi}{2}$ is needed to maintain the unitary pairing.
 
In summary, we have investigated the competing pairing symmetries
in  ultracold multicomponent dipolar molecular systems,
which provides a wonderful opportunity to investigated exotic
pairings. 
We predict that  the anisotropy of the dipolar interaction 
provides a well-defined pairing mechanism to the 
spin triplet, or, more generally, the spin symmetric 
channel Cooper pairing.
The spin singlet even parity channel pairing in the $\phi^{s+d}$ channel is 
tunable by the short range $s$-wave scattering.
It mixes with the spin triplet odd parity channel pairing by developing
a relative phase $\pm\frac{\pi}{2}$ to maintain the unitary pairing.
This is another type of unconventional Cooper pairing breaking
TR symmetry.

We point out that our mechanism of TR breaking mixing between triplet
and singlet pairings is very general.
For example, in superfluid $^3$He it was
originally proposed that pairing would occur in the singlet $d$-wave channel
\cite{dwave}, induced by the attractive
part of the Van der Waals interaction. 
Later, attention focused exclusively  on pairing in  the
triplet $p$-wave   channel   induced by the short-range repulsive 
interaction  \cite{dwave,abm}.
It is natural to expect that both channels
could contribute to pairing at sufficiently low temperatures,
leading to coupled Balian-Werthamer (BW) triplet pairing
and singlet channel pairing with a relative phase $\pm \frac{\pi}{2}$.
In metallic superconductors, coupling of an isotropic $s$-wave state with a 
BW triplet state will lead to a single isotropic gap only 
for the particular case where the relative phase is  $\pm\frac{\pi}{2}$.
These possibilities will be discussed separately.

C. W. thanks helpful discussions with E. Fradkin and S. Kivelson.
C. W. is supported by NSF under No. DMR-0804775, and Sloan Research
Foundation. This work was also supported in part by 
NSF under No. PHY05-51164 at the KITP.

\end{document}